\documentstyle[11pt,epsfig]{article}
\columnwidth\textwidth
\begin{document}
\begin{flushright}
BUTP-9518 \\
hep-ph/9505209
\end{flushright}

\begin{center}
\Large{ \bf An exact analytic description
of neutrino oscillations in matter with exponentially varying density
for arbitrary number of neutrino species}
\end{center}

\begin{center}
E. Torrente Lujan.
Inst. fur Theoretische Physik, Universitat Bern \\
Sidlerstrasse 5, 3012 Bern, Switzerland.\\
e-mail: e.torrente@cern.ch\\
\end{center}

\begin{abstract}

Exact analytical expressions in terms of generalized confluent hypergeometric
functions for the transition amplitudes of 
neutrino oscillations in 
presence of matter are computed for an arbitrary number
of species.
The density of matter is assumed to be exponentially decaying. The 
results can be used for the 
description of matter-induced neutrino oscillations in the Sun which
can take place when the solar neutrinos propagate radially
from the interior to the 
surface.  
Expressions are particularly simple in the limit of infinite
propagation time as is suitable for the case of detection at Earth.


\end{abstract}

\newpage

\section{Introduction}

The explanation of the solar neutrino flux problem by maximally mixed neutrino
oscillations in vacuum or by resonance-enhanced oscillations without large
mixing (MSW effect) is well known by now and it has been treated 
abundantly in
the literature (\cite{panta1,mik1,pet1}).

For stable and relativistic 
neutrinos $\nu=(\nu_i, i=1,N_\nu)$ of
equal momentum the propagation in vacuum is described 
in the basis of mass eigenstates by
$$i\partial_t\nu=H^0\nu$$
where
\begin{equation}
H^0=Diag(E_i=\surd (p^2+m^2)\simeq p+m_i^2/2E)
\label{e02}
\end{equation}
The common constant $p$ can be absorbed as an unimportant phase and is
usually omitted.

The weak eigenstates $\nu'$ are related to mass eigenstates by
($u^\ast$ for antineutrinos)
$$\nu'=u \ \nu$$
The transitions between $\nu'$ 
are described by the amplitude matrix
$${\cal A}=ue^{-i H^0 t} u^{-1} $$

In presence of standard matter with arbitrary electron number density,
the propagation is usually well aproximated by 

\begin{equation}
i\partial_t\nu=(H^0+\rho(t) u A u^{-1})\nu
\label{e50}
\end{equation}
where A is a matrix with $A_{11}$ as only non-zero element.
$\rho(t)$, essentially a forward scattering amplitude, is proportional to the 
electron number density of the medium
$$\rho(t)=\pm\surd2 G_F N_e(t)$$
The plus/minus sign is for neutrinos/antineutrinos
respectively.

The difference of the eigenvalues of $H^0$ (
the inverse of the usually defined oscillation length)
is typically considered to be, in 
the solar case, of the
order
$$\frac{m_i^2-m_j^2}{2 E}\approx \frac{0.1-1\ eV^2}{ 1\ MeV}\approx 10^{-6}-10^{-7}\
eV $$ 

The electron number density can be parametrized (data taken from \cite{bah1}) , for sufficiently far distances
from the solar core as  
$$N_e(r)= N_0 \exp(-\lambda r)=160\pm 30 \exp (-10.6\pm 0.2\  r/r_0)\  mol/cm^3 $$
with $r,r_0$ the distance from the center and solar radius respectively.
At the solar center 
$$N_e(0)\approx 100 \ mol/cm^3 $$

So the  value for $\rho(t)$  varies between 
zero and its maximum value of $\approx 10^{-11}$ eV at the center.

Later we will see that the following dimensionless quantities play an important role:
$$\frac{m_i^2-m_j^2}{2 E\lambda}\approx 1.75\ 10^8 
\frac{\Delta m^2\ (eV^2)}{E\ (Mev)}
\; \; (\approx 1\,  \hbox{  for E= 1 MeV and $\Delta m^2=10^{-7}\ eV^2$)} $$

$\rho(t)/\lambda$ varies between  $5.5\ 10^3$ and $\approx 0$  when the neutrino goes
outwards from the center to the sun surface.

In the general case  Eq.(\ref{e50}) must be solved numerically 
(\cite{mik1,pet1}).
Comparation with  numerical studies and qualitative analyses show 
that for slow (in the neutrino 
time scale ) and monotonically varying matter density, the adiabatic
approximation can be applied except, possibly, in some small ''resonance''
region (MSW effect). 
This resonance effect is essentially 
the same as the non-adiabatic crossing of energy levels in diatomic molecules
studied already  60 years ago.
The overall transition probability depends  essentially 
of the existence of this resonance region and not on the detailed 
matter density function.  The  aproximate transition probability was
computed 
for two species in two classical papers  by Landau and Zener
\cite{zen1} supposing that the density varies linearly
in the (small) resonance region, its formula, adapted to the
neutrino context, is:

\begin{equation}
P^{LZ}_{ee}=\exp\left [ -\pi \frac{\Delta m^2}{2 E\lambda}\frac{s^2 2\theta}{c 2\theta}\right ]
\end{equation}

The point $t_{res}$ where the resonance  appears
is given by
$$\rho(t_{res})=\frac{\Delta m^2 \cos 2\theta}{E} $$
$\theta$ is the 2-neutrino mixing angle.


Apart from this approximate results, exact solutions have appeared for
particular forms of the function $\rho $:  for linear densities,
in terms of Weber-Hermite functions (\cite{pet2,hax1}); for functions
of the form
$$\rho(t)= C (1+\tanh (\lambda t)) $$
in \cite{not1}, and for exponentially decaying densities
$$\rho(t)= c e^{-\lambda t} $$
in \cite{pet3,tov1}.
All these functions for the density are of interest in the solar case, the last
two reproduce rather well the real solar density except in the inner
core and to a less degree in the surface boundary. 
All of these solutions are valid for 2 species and all of them are obtained
reducing Eq.(\ref{e50}) to a second order differential equation identifiable with
one of the classical equations of the mathematical physics.

In this paper , using a completely different method, 
summing a standard time-dependent
perturbative expansion, we will give an exact, analytical solution for an arbitrary number
of species and for arbitrary (possibly non-unitary in case of sterile neutrinos)
mixing matrix using the exponentially decaying function. Using the limit
$t\to\infty$, which is appropriated for the case of 
detection at Earth, we will give specially simple solutions (this is
something that papers \cite{pet3,tov1} fail to give in the 2-dimensional case).

\section{An exact solution for $\delta(t)\approx \exp (-t)$.}

We will  get an exact solution for the neutrino evolution equations in 
the special case where the electron density decays exponentially.
Standard non-stationary perturbation theory will be used;  we will give general
expressions for the n-th term and sum the full series (regardless of mathematical
convergence problems, we will see a posteriori that the series effectively
converges, and the result is physically meaningful).

The evolution operator of the differential system 

\begin{equation}
i\partial_t \nu= H(t)\nu 
\label{e201}
\end{equation}
is such that

\begin{equation}
 \nu(t)=U(t,t_0)\nu(t_0)
\end{equation}
obeying the integral equation

\begin{equation}
U(t,t_0)=1-i\int_{t0}^t H(\tau)U(\tau,t_0)d\tau
\label{e102}
\end{equation}

We are interested in the case where the hamiltonian, of finite dimension k , can be decomposed in the
following way

\begin{equation}
H=H^0+V(t)=H^0+ \rho(t) V
\end{equation}
$H^0$ ,the free hamiltonian, hermitic and independent of time, has eigenvalues $\{E_1,\dots, E_k \}$ and eigenvectors
$\{ \mid a >,\mid b >\dots \}$.
We can suppose that at least two of the eigenvalues are distinct, the completely
degenerate case can be solved in a trivial way.
$\rho(t)$ a scalar 
function and V a hermitic matrix with all eigenvalues but one
equal to zero.

 Formally, it is 
possible to solve  Eq.(\ref{e102}) by succesive iterations:

\begin{equation}
U(t,t_0)=U^{(0)}(t,t_0)+\sum_{n=1}^{\infty}U^{(n)}(t,t_0)
\end{equation}

where the order 0 can be taken as:

\begin{equation}
U^{(0)}(t,t_0)=\exp\left (-i H^{0}(t-t_0)\right )
\end{equation}
 and $U^{(n)}$ is the well-known integral

\begin{equation}
U^{(n)}=(-i)^n\int_{\Gamma}
d\tau_n
\dots d\tau_1 
U^0(t,\tau_n)V(\tau_n)
\dots
U^0(\tau_2,\tau_1)V(\tau_1)U^0(\tau_1,t_0)
\label{e101}
\end{equation}

The domain of integration is defined by 
$$\Gamma\equiv t>\tau_n>\dots >\tau_1>t_0.$$
A V with the special form of interest to us
can be always reduced to the form

\begin{equation}
V= u^{-1} 
\ Diag (1,0,\dots,0) \ u 
\label{e100}
\end{equation} 
for some unitary  matrix u and
its elements are given   by

\begin{equation}
V_{ij}=\sum_{k,l} u_{ik}^{-1}\delta_{k1}\delta_{l1}
u_{lj}^{\phantom{-1}}= u_{1j}^{\phantom{-1}} u_{i1}^{-1}
\label{e120}
\end{equation}

The following property will be used later valid for any V of this form: 
$$V_{ij} V_{jl}= V_{il} V_{jj} \quad \hbox{(no summation involved)}$$

 First we will compute 
U in the $t\rightarrow\infty$ limit. 
And the end we will see that in fact we can compute it for any finite t
using the properties of the 
evolution operator.
For simplicity, we will work in the basis of eigenstates of
$H^0$.
For practical applications (weak eigenstates neutrino transition
probabilities) we may want to apply  a straightforward
coordinate change $ U'= v^{-1} U v $ later.

 Trought elementary manipulations we get, from Eq.(\ref{e101}),
\begin{eqnarray}
\lefteqn{<b\mid U^{(n)}\mid a> =  
(-i)^n\exp(-i (E_b t-E_a t_0)) \times}\nonumber \\
 & &\times \sum_{k1,\dots,k(n-1)}\int_{\Gamma}
d^n\tau
\exp(i\tau_n w_{bk1}+\dots+ i\tau_1 w_{k(n-1)a} ) V_{bk1}(\tau_n)\dots
V_{k(n-1)a}(\tau_1) 
\end{eqnarray}

With $w_{k1k2}=E_{k1}-E_{k2}$.
Besides, using the special  form of the potential:

\begin{equation}
\begin{array}{c}
<b\mid U^{(n)}\mid a>=(-i)^n\exp (-i (E_b t-E_a t_0)) 
V_{ba}\  
\sum_{k1,\dots,k(n-1)} B_{k1}\dots B_{k(n-1)} A_{k1,\dots,k(n-1)}\\ \\
A_{k1,\dots,k(n-1)}=
\int_\Gamma d^n\tau
\exp( i\tau_n w_{bk1} +\dots +i\tau_1 w_{k(n-1)a})
\rho(\tau_n)\dots \rho(\tau_1)\\ \\
B_k=V_{kk}
\end{array}
\end{equation}
The multiple sums run over all the eigenvalues of the hamiltonian
independently.

We  consider an idealized exponentially decaying electron density profile
for the sun: 
$$\rho(t)=\rho_0 \exp(-\lambda t)$$

Where $\rho_0$ can be positive or negative. 
Redefining $E_k,\rho_0$ we can always suppose $\lambda=1$.

For this kind of density we can use the
following equality,
\begin{eqnarray}
I_n( w_1,\dots,w_n) &
\equiv &
\int_{t0}^\infty\dots\int_{t0}^{x_{ n-1}} dx_1 \dots dx_n\exp
\sum_n w_n x_n \nonumber\\
& =&
   \frac{ (-1)^n \exp (t_0\sum_n w_n)}{w_n (w_n+w_{n-1})\dots (w_n+w_{n-1}\dots +w_1)}\nonumber \\
& & \nonumber\\
& & (\hbox{valid if\ }\ \  \Re{\ w_n} < 0,\ \forall\ n\ ) \label{e612}  
\end{eqnarray}

This equality can be proved by induction, noting that, with the
 help of the variable change 
$$ x_1=y_1+y_2 ;\quad  x_2 = y_2   $$ 

 the following recurrence relation holds
$$ I_n(w_1,\dots,w_n)=I_1(w_1)\times I_{n-1}(w_1+w_2,w_3,\dots,w_n)$$
that together with
$$I_1(w)=\frac{- \exp (w\ t_0)}{w}$$
proves our result.

In our case 

\begin{equation}
\begin{array}{rl}
w_n+w_{n-1}+\dots + w_{n-j+1} &= i w_{bk1}-1+ i w_{k1k2}-1+\dots +i
w_{k(j-1)kj}-1\\
 &= iw_{bkj}-j
\end{array}
\end{equation}

so the (n-1)-tensor $A_{i..j}$ can be factorized 

\begin{equation}
A_{k1,k2,\dots,k(n-1)}=e^{iw_{ba}t_0}\frac{\left(-\rho_0 e^{-t_0}
 \right)^n}{i w_{ba}- n }A_{k1}A_{k2}
\dots A_{k(n-1)}
\end{equation}
with

\begin{equation}
A_{k(m)}=\frac{1}{iw_{bk(m)}- m}
\end{equation}
 the fact 
that this factorization is possible is the key for the solution of the 
problem.
Note also the importance of the ordering in the definition
of the A's.

So,

\begin{equation}
\sum B_{k1}\dots B_{k(n-1)} A_{k1}\dots A_{k(n-1)}\propto
\prod_{m=1}^{n-1} \sum_{all\ eigenvalues} B_{k(m)} A_{k(m)}\equiv\prod_m f_m
\end{equation}

What remains is the computation of the expression $\prod_m f_m$.
Such computation is easy even in a general case but rather messy. 
We will do it first for  the 2-dimensional case for the sake of clarity.

\subsubsection{2-dimensional case}

For a two-dimensional case and taking b=1

\begin{equation}
f_m=\frac{B_1}{-m}+\frac{B_2}{i w_{12}-m}
=\frac{-1}{m} \frac{\left ( -iB_1  w_{12} + m\right )} 
{\left ( -iw_{12} +m \right )}
\label{e610}
\end{equation}
and the product of them

\begin{equation}
\prod_{m=1}^{n-1} f_m=\frac{1}{B_1}\frac{(-1)^{n-1}}{ 
(n-1)!}\frac{( B_1 \beta)_{(n)}}{ (\beta)_{(n)}}
\end{equation}
where we have used the Pochammer symbol Eq.~(\ref{a110})
 and
defined $\beta=-iw_{12}$.
Inserting this product in the previous expressions and recalling the
series expansion of the Confluent 
Hypergeometric function Eq.~(\ref{a100})
we can write immediately the value  of the 
diagonal elements of U 

\begin{eqnarray}
\lefteqn{<b\mid U\mid b>=} \nonumber \\
 &= & e^{-i E_b (t-t_0)} \left ( 1+V_{bb}\sum_{n=1} (-i)^n \frac{\left ( -\rho_0
\exp(-t_0)\right )^n}{-n}\prod_{m=1}^{n-1} f_m \right ) \nonumber \\
 &= & e^{-i E_b (t-t_0)}
\left ( 1+ \frac{V_{bb}}{B_b}\sum_{n=1} \frac{\left (-i\rho_0\exp(-t_0) \right)^n}{n!}
\frac{(B_b\beta)_{(n)}}{\beta_{(n)}}\right ) \nonumber \\
&=& e^{-i E_b (t-t_0)}\  \  \mbox{${{}_1 F_1}$} (B_b\ \beta, \beta, z)
\end{eqnarray}
here $\beta=-iw_{ba}$ and $z=-i\rho_0\exp(-t_0)$.

For the general case $a\not = b$ is neccesary to work a little bit more to give a
closed expression. In this case it appears an additional factor
in each term of the series which doesn't allow for the immediate 
identification
with any known function:

\begin{eqnarray}
\lefteqn{<b\mid U\mid a>=} \nonumber \\
 &= & e^{-i (E_b t- E_a t_0)} \left ( \delta_{ab}+e^{i(E_b-E_a)t_0} V_{ba}\sum_{n=1}
(-i)^n \frac{\left ( -\rho_0
\exp(-t_0)\right )^n}{-\beta-n}\prod_{m=1}^{n-1} f_m \right ) \nonumber \\
 &= & e^{-i (E_b t- E_a t_0)}
e^{i (E_b-E_a) t_0} 
 \frac{V_{ba}}{B_b } 
\sum_n \frac{z^n}{(n-1)!}
\frac{1}{\beta+n} \frac{(B_b\beta)_{(n)}}{\beta_{(n)}} \nonumber \\
& = & e^{-iE_b(t-t_0)}\frac{V_{ba}}{B_b} 
g(B_b \beta, \beta; z)
\end{eqnarray}
as before here $\beta=-iw_{ba}$.

 The function
\begin{equation}
 { g}(z)=\sum_{n=1}^\infty
\frac{1}{\beta+n}\frac{(\alpha\beta)_{(n)}}{(\beta)_{(n)}} \frac{z^n}{(n-1)!} 
\end{equation}
is well defined, the series
is absolutely convergent for all z; manipulating termwise, and 
using the Eq.(\ref{a120})
we get the following
 differential equation for it 

\begin{equation}
z g' +\beta  g= \alpha z\ \mbox{${{}_1 F_1}$}(1+\alpha\beta,1+\beta;z);\quad
\quad g(0)=0
\end{equation}

whose solution is

\begin{equation}
g(z)=\alpha z^{-\beta}
\int_0^z y^\beta \mbox{${{}_1 F_1}$}(1+\alpha\beta,1+\beta;y)dy
\end{equation}

Using the Expression (\ref{a130}) 
for the special case $c=\gamma+1$, we finally write (note that in our
case the conditions of validity of the previous integral are fullfilled) :
\begin{eqnarray}
g(z)& =&\frac{\alpha}{1+\beta}z\ \mbox{${{}_1 F_1}$}(1+\alpha\beta,2+\beta;z)
\nonumber \\
&=& \mbox{${{}_1 F_1}$}(\alpha\beta,\beta;z)-\mbox{${{}_1 F_1}$}(\alpha\beta,1+\beta;z)
\label{e158}
\end{eqnarray}

The remarkable second equality can be deduced from the expressions
(\ref{a120}-\ref{a160}).
So
\begin{eqnarray}
\lefteqn{<b\mid U\mid a>=} \nonumber \\
 &= & e^{-i E_b (t-  t_0)} 
 \frac{V_{ba}}{B_b } \frac{B_b}{1+\beta}
\ z\ \mbox{${{}_1 F_1}$}(1+B_b \beta, 2+\beta; z)  \nonumber \\
 & & \nonumber \\
\lefteqn{<a\mid U\mid b>=} \nonumber \\
 &= & e^{-i E_a (t-  t_0)} 
 \frac{V_{ab}}{B_a } \frac{B_a}{1-\beta}
\ z\ \mbox{${{}_1 F_1}$}(1-B_a \beta, 2-\beta; z)  \nonumber \\
 &=& e^{-i E_a (t-  t_0)} 
  \frac{V_{ab}}{1-\beta}
\ z\ e^z \mbox{${{}_1 F_1}$}(1-(1-B_a) \beta, 2-\beta; -z)  \nonumber \\
 &=& e^{-i E_a (t-  t_0)} 
  \frac{V_{ab}}{1-\beta}
\ z\ e^z \mbox{${{}_1 F_1}$}(1-B_b \beta, 2-\beta; -z) \nonumber  \\
 &=& e^{-i E_a (t-  t_0)} 
  \frac{V_{ab}}{1-\beta}
\ z\ e^z \mbox{${{}_1 F_1}$}^\ast(1+B_b \beta, 2+\beta; -z^\ast) \label{e1502} \\
&=& e^{-i E_a (t-  t_0)} 
  \frac{V_{ab}}{1-\beta}
\ z\ e^z \mbox{${{}_1 F_1}$}^\ast(1+B_b \beta, 2+\beta; z)  \label{e1501}
\end{eqnarray}

Where we have used the Eq.(\ref{a120}). The last Formula (\ref{e1501}) is
only valid for $z$ purely imaginary, in the case of general complex $z$ (for
example given $\rho_0$ complex ) Formula (\ref{e1502}) is valid. 

We have completed so the computation of U.
In summary, the elements of the operator U in a basis of eigenvectors
of $H^0$ are 
\begin{eqnarray}
U(t,t_0)&=&\exp-i H_0 (t-t_0) U_{red}( \rho_0,t_0)\nonumber \\
     & & \nonumber \\   
U_{red}(\rho_0,t_0)&=&\pmatrix{
 F &  \frac{V_{12}}{V_{11}}  \  G \cr
 -\frac{V_{21}}{V_{11}}
\ \exp(z) \ G^\ast& \ \exp z \ F^\ast \cr }
\label{e150}
\end{eqnarray}
with the shorthands
$$ G=g(z)= \frac{V_{11}}{1+\beta}\ \mbox{${{}_1 F_1}$} (1+B_1\ \beta, 2+\beta;z) 
,\quad F=\mbox{${{}_1 F_1}$} (B_1\ \beta, \beta;z) $$

As it should be, $U(t)$ obeys the free equation:
$$i\partial_t U(t)= H^0\ U(t) $$

For a reverse sign in $\rho_0$ (antineutrino case), the matrix $U_{red}$ is
basically the same except for the substitution $B_1\to 1-B_1$:
\begin{eqnarray}
U_{red}(-\rho_0,t_0) &=&\exp(-z)\pmatrix{
 F^\ast &  -\frac{V_{12}}{V_{11}}  \  G^\ast \cr
 \frac{V_{21}}{V_{11}}
\ \exp(z) \ G& \ \exp z \ F \cr }
\label{e151}
\end{eqnarray}
where now F stands for $\mbox{${{}_1 F_1}$}((1-B_1)\beta,\beta,z)$, and G changes similarly.

For H hermitic  U is unitary and we get as a byproduct
 the following non-trivial identity for the
absolute value of Hypergeometric functions:

\begin{equation}
\left \| \mbox{${{}_1 F_1}$}(\alpha \beta,\beta; z)\right \|^2+\alpha
(1-\alpha)\left \|\frac{z}{1+\beta}\mbox{${{}_1 F_1}$}(1+\alpha\beta,2+\beta;z)\right \|^2\equiv 1
\quad \hbox{($\alpha$ real)}
\label{e202b}
\end{equation}  

or, using the second Equality in (\ref{e158})

\begin{equation}
\mid \mbox{${{}_1 F_1}$}(\alpha \beta,\beta; z)\mid^2
+\left ( \frac{1-\alpha}{\alpha}\right )\mid\mbox{${{}_1 F_1}$}(\alpha\beta,1+\beta;z)- \mbox{${{}_1 F_1}$}(\alpha \beta,\beta; z)\mid^2\equiv 1
\quad \hbox{($\alpha$ real)}
\label{e202}
\end{equation}

For the particular case of  u being the 
orthogonal 2-dimensional matrix 
$$
u=\pmatrix{C\theta & -S\theta \cr
           S\theta &  C\theta  \cr }
$$
the matrix V is
$$
V=\pmatrix{C^2\theta & -S\theta C\theta \cr
           -S\theta C\theta &  S^2\theta  \cr }
$$
and $H^0$ the standard neutrino
oscillation hamiltonian  given by Eq.(\ref{e02}). 
We arrive to ($t_0=0$ and restoring $\lambda$),
\begin{eqnarray}
U_{11}^{red}&=& 
\mbox{${{}_1 F_1}$}( \frac{iw_{21}}{\lambda} C^2\theta, \frac{iw_{21}}{\lambda}; 
\frac{-i\rho_0 }{\lambda} )\label{e611b} \\
U_{12}^{red}&=&\  T\theta\  \frac{-i\rho_0/ 
\lambda}{1+iw_{21}/\lambda} 
\ \mbox{${{}_1 F_1}$}\left (1+ \frac{iw_{21}}{\lambda}\ C^2\theta, 2+\frac{iw_{21}}{\lambda}; 
\frac{-i\rho_0 }{\lambda} \right ) \label{e611}
\end{eqnarray}

We are interested in the expression of the evolution operator  in the basis of eigenvalues
of V,  it is straigthforward to compute $U'=u U u^{-1}$.
After some algebra, applying the identities we have seen before
and averaging out time dependent terms, the probability of electron 
permanence is
\begin{eqnarray}
P_{ee}=\left\| U_{11}'\right\|^2&=&1- S^2\theta \left (1+ C 2\theta \left \| \mbox{${{}_1 F_1}$}\left ( \frac{iw_{21}}{\lambda} C^2\theta, 1+\frac{iw_{21}}{\lambda}; 
\frac{-i\rho_0 }{\lambda}\right ) \right \| ^2\right ) 
\label{e550}
\label{e203}
\end{eqnarray}

An expression equivalent to the previous one  has been obtained already by
\cite{aba1} in terms of Whittaker functions. However our expression is simpler
and more compact because we use the infinite time limit.

In the limit $\rho_0\to 0$  or  free case, $z\to 0$ and
the hypergeometric functions go to 1, U becomes as expected:
$$ U=\exp -i H^0 t $$

In the limit of vanishing mixing angle, $C^2\theta\to 1$, $\tan\theta\to 0$ and
$\mbox{${{}_1 F_1}$}(z)\to\exp z$ in (\ref{e611b}), so
\begin{eqnarray}
P_{ee}&=&1 
\end{eqnarray}
In the limit $w_{21}/\lambda=\Delta m^2/2 E\lambda\to 0$, $\mbox{${{}_1 F_1}$}\to 1$ in
(\ref{e550}) and 
\begin{eqnarray}
P_{ee}&=&1-\frac{1}{2}S^2 2\theta
\label{e810}
\end{eqnarray}

 In the 
 the limit $w_{21}/\lambda>> 1 $ (adiabatic regime) we 
distinguish two cases: if $ w_{21}/ \lambda\not\simeq \mid z\mid $ we have 
 $\mbox{${{}_1 F_1}$}\to \exp
(-i C^2\theta \rho_0/\lambda)$ and Formula~(\ref{e810}) also applies;
if $w_{21}/\lambda\simeq \mid z\mid$ a resonance occurs and the probability
drops abruptly. Some asymptotic formulas exist for this case valid except in the
central region of the resonance (see \cite{grad,din1}). 

In the limit of small (but not vanishing) mixing angle, we are
left in principle with the  expression for the electron permanence probability
(computed ignoring diagonal terms in U)
\begin{eqnarray}
P_{ee}=&(1-2 C^2\theta S^2\theta) \left \|
\mbox{${{}_1 F_1}$}\left ( \frac{iw_{21}}{\lambda} C^2\theta, \frac{iw_{21}}{\lambda}; 
\frac{-i\rho_0 }{\lambda}\right ) \right\|^2 \label{e560}
\end{eqnarray}

In the figures we show a comparison between 
formulas (\ref{e550}) and (\ref{e560}). We see that even at very small angles
Eq.(\ref{e560}) 
is not a very good approximation of (\ref{e550}) as we could expect.

\begin{figure}[p]
\centering\hspace{0.8cm}
{\epsfig{file=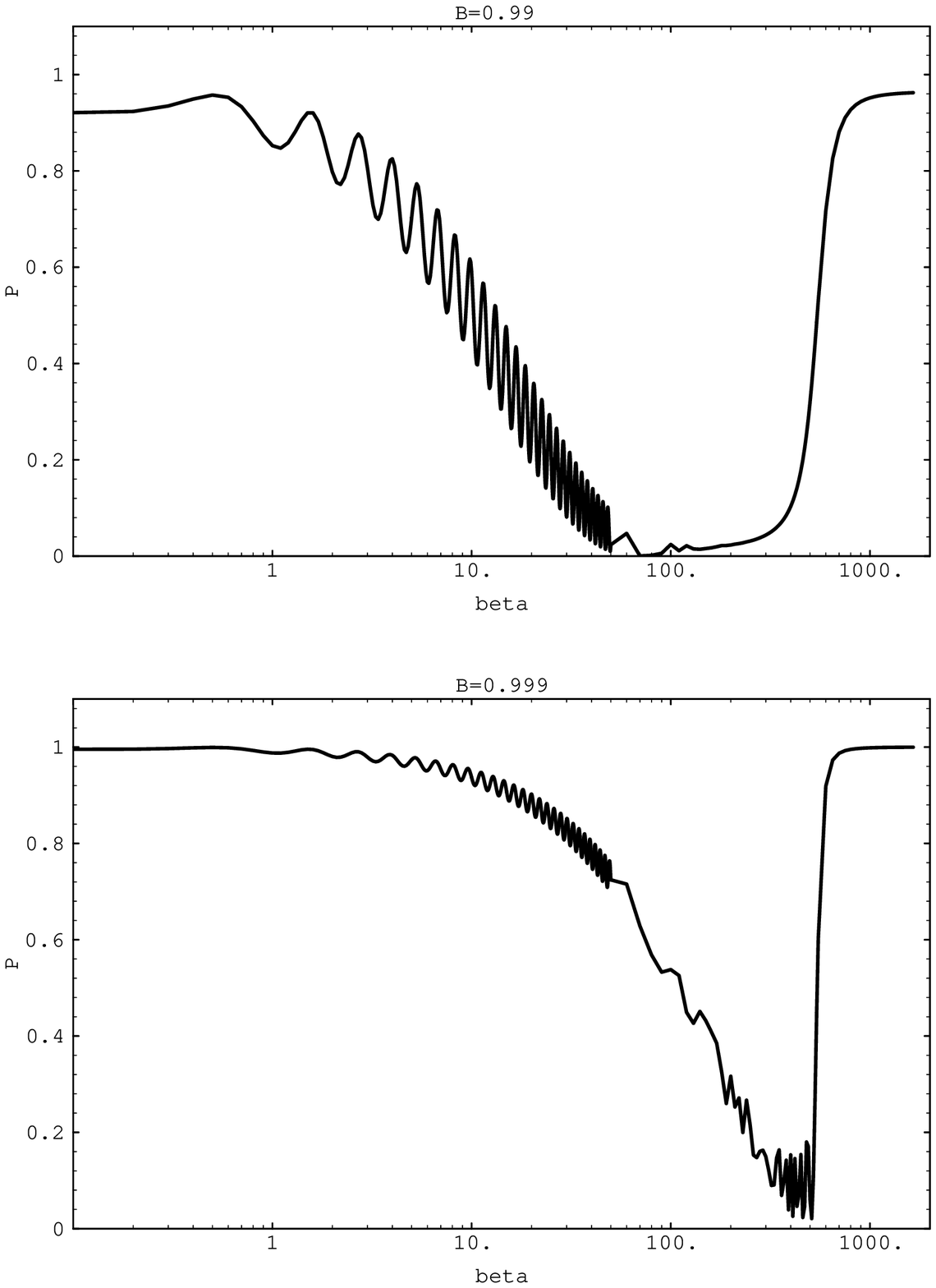,height=12cm}}
\caption{The function 
\protect{$\mid \mbox{${{}_1 F_1}$}( B i \beta, i \beta, -i 550.2)\mid^2$} 
for two different B's near 1, as a function of \protect{$\beta$}.
This corresponds approximately to the oscillation probability for a neutrino
produced at $r/r_0=0.25$ 
as function 
of \protect{$\beta=\Delta m^2/2E\lambda$}.}
\label{f1}
\end{figure}

In Figure (\ref{f1})  we plot 
$\mid\mbox{${{}_1 F_1}$}\mid^2$ (Eq.(\ref{e560})) as a function of its parameters keeping z 
constant.  Increasing values of the x-axis implies  a larger mass 
difference or a smaller neutrino energy.
We observe that the probability is near 1 
( adiabatic evolution ) except for a very concrete region of
the parameter space (MSW effect or existence of  a resonance layer).
The starting point and width of this region (but not its end point) depends on
the parameters involved, in particular $\cos^2\theta$.

\begin{figure}[p]
\centering\hspace{0.8cm}
{\epsfig{file=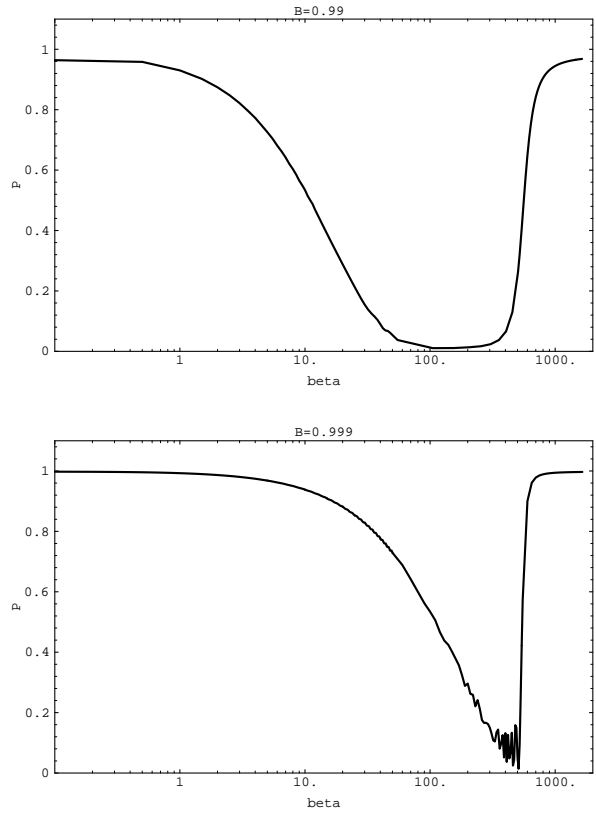,height=12cm}}
\caption{$P_{ee}$ (two neutrino species, Formula~\protect{\ref{e203}}) for 
a neutrino produced at $r/r_0=0.25$ 
as a function of
$\beta=\Delta m^2/2E\lambda$ and two different mixing angles: 
$B(=\cos^2\theta)= 0.99, \ 0.999$.}
\label{f2}
\end{figure}

\begin{figure}[p]
\centering\hspace{0.8cm}
{\epsfig{file=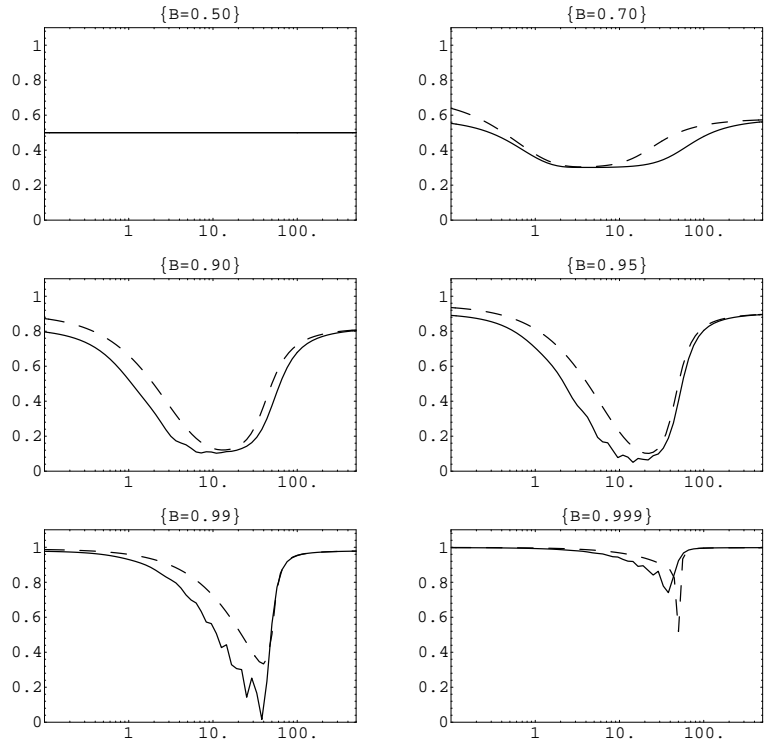,height=15cm}}
\caption{
As in  Figure~(\protect{\ref{f2}}) but for a neutrino created
near the surface ($\mid z\mid\approx 20$) as a function of the mixing angle. The
dashed line corresponds to the Parker approximate solution (as it is given by
Formula 2.18 in \protect\cite{panta2}).}
\label{f3}
\end{figure}

\begin{figure}[p]
\centering\hspace{0.8cm}
{\epsfig{file=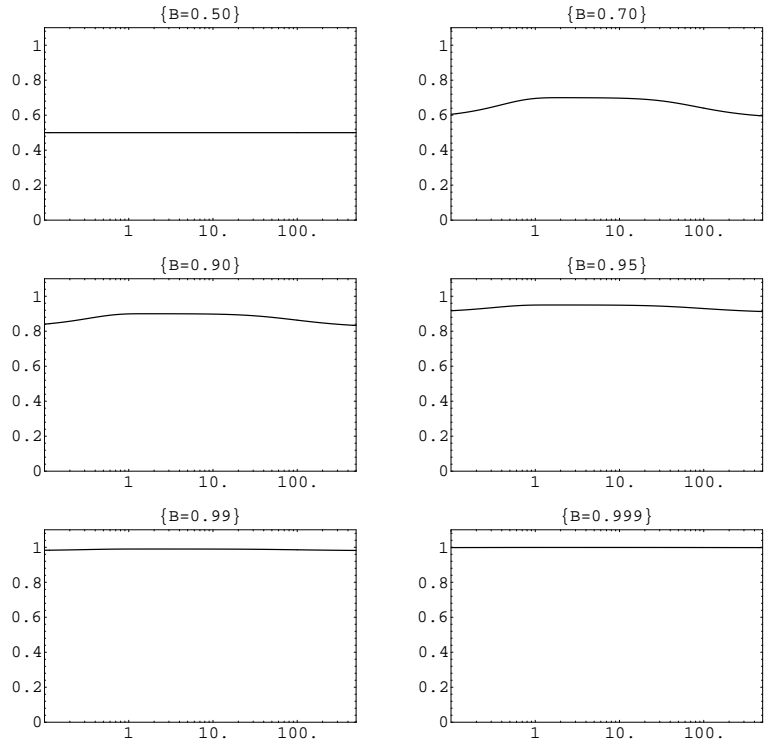,height=15cm}}
\caption{
As Fig.~(\protect{\ref{f3}}) but for antineutrinos (reverse sign in the argument
$z$ of the hypergeometric function).In this case there is not resonance (it is
possible to distinguish some ''anti''-resonace behaviour for larger mxing angles).}
\label{f5}
\end{figure}

Figure (\ref{f2})  is equivalent to Fig.(\ref{f1}), this time we plot
the exact expression for the e-e probability, Eq.(\ref{e550}), 
The global behaviour is exactly the same as before, the local oscillations
have disappeared now however.  
Figure (\ref{f3}) show the 
behaviour of the resonance region also for large mixing angles.
The continuos curve is our exact solution, the dashed curve the Parker
approximate formula as given in (\cite{panta2}).
 Figure (\ref{f5}) show the behaviour
for antineutrinos (reverse sign for $\rho_0$ in z), in this case the resonance
is absent.

\begin{figure}[p]
\centering\hspace{0.8cm}
{\epsfig{file=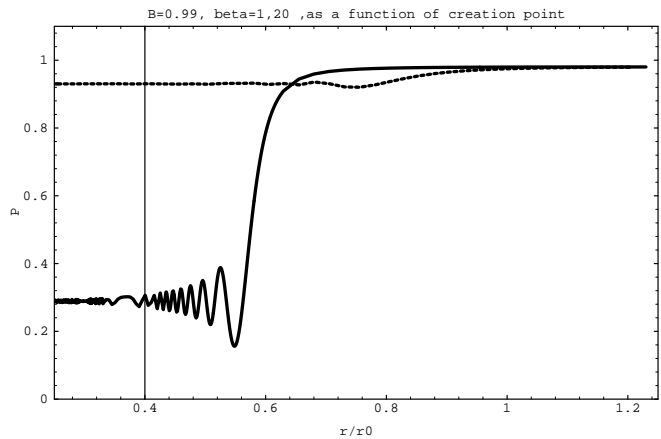,height=12cm}}
\caption{
$P_{ee}$ (Eq.~(\protect{\ref{e203}})),
as a function of the creation point (in solar radius fraction) for a fixed
mixing angle $\cos^2\theta=0.99$ and two different $\beta=\Delta
m^2/2E\lambda=1,20$ (respectively upper and lower curves ).
}
\label{f6}
\end{figure}

\begin{figure}[p]
\centering\hspace{0.8cm}
{\epsfig{file=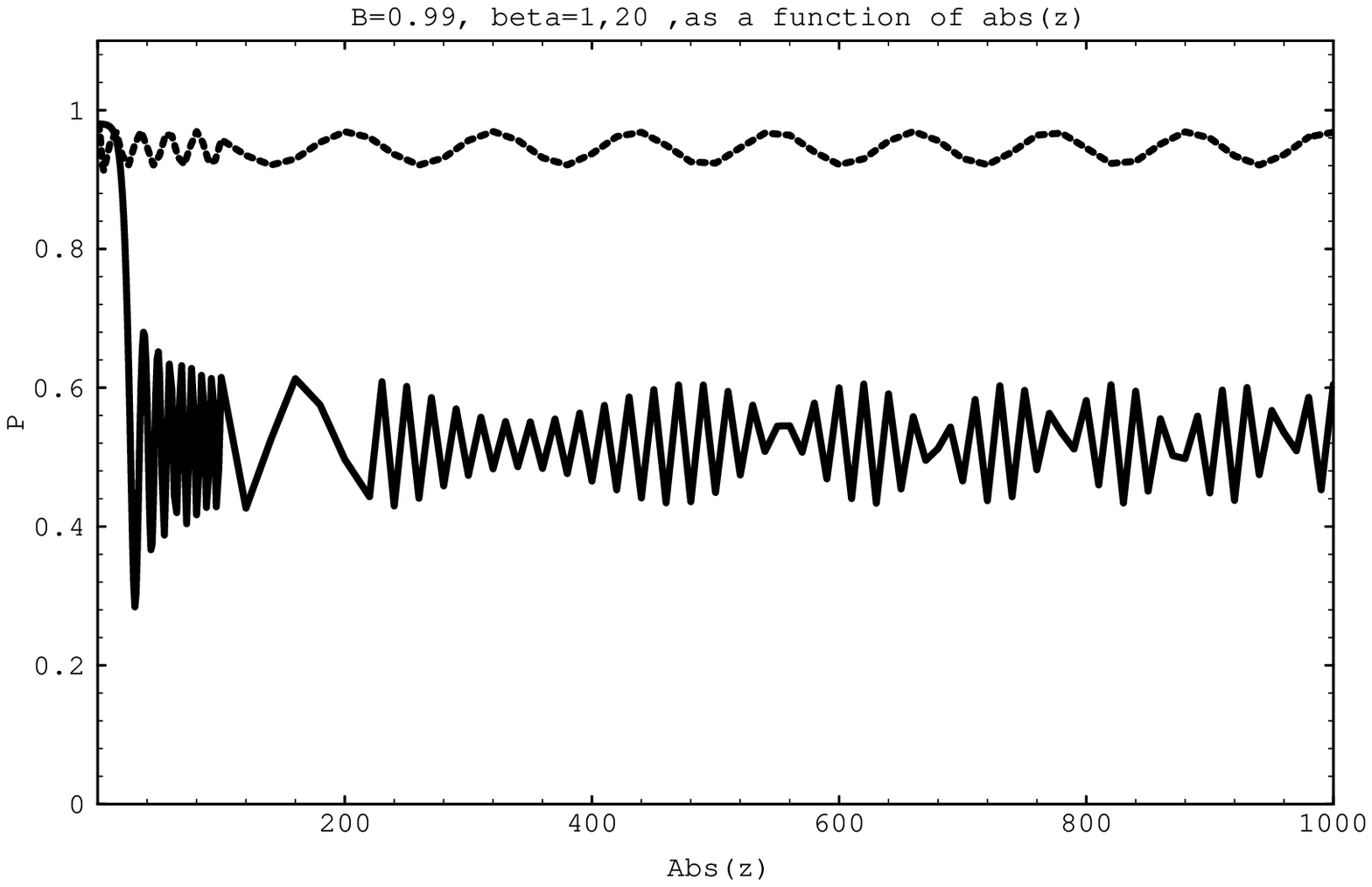,height=12cm}}
\caption{
 Function $\mid \mbox{${{}_1 F_1}$} (i B \beta,i \beta,z)\mid^2 $ as a
function of $\mid z\mid$, for different values of the parameters. B=0.99, the upper and
lower curves correspond respectively to $\beta=$ 1 and 20.}
\label{f7}
\end{figure}

In a second set of plots (figs.(\ref{f6},\ref{f7})) we keep constant the parameters and vary
 z or equivalently the production point of the neutrino: larger $\mid z\mid$ 
or $\rho_0$ means 
larger density at the creation point which implies that this one
is nearer to the center. We see again there is  a transition between
two well defined zones: for a very far creation point, the neutrino
doesn't pass through any resonance layer, adiabatic regime conditions 
always apply and $P_{ee}\sim 1$.

In order to study in detail the existence  and properties of the resonance
region it would be convenient to have general formulas for the zeros of the
Hypergeometric functions, as function not only of its argument but also of its
parameters. Unfortunely very little is known in a general case.
There are some results (see appendix and
reference therein) about the real zeros of $\mbox{${{}_1 F_1}$}(a,b;x)$ with $a,b$ real.

We note that the expressions (\ref{e150}) and (\ref{e560})
are still formally valid for a purely imaginary $\lambda$ (we take for granted
that we can take convenient limits $\Re w_n\to 0$ in the integral (\ref{e612})).
Expression (\ref{e550}) however is not valid any more  because we used the
hermiticity of H (and unitarity of U) in computing it. 
The extrema of  $\mid\mbox{${{}_1 F_1}$}\mid^2$ are given essentially by the zeros of 
$\mbox{${{}_1 F_1}$}(1+a,1+b;x_j)$;
for $\lambda$ purely imaginary, argument and parameters in \mbox{${{}_1 F_1}$}\ become real and we can apply the bounds in Appendix A
given by the Eqs.(\ref{a170}-\ref{a190})\ .
This bounds can be used to deduce the regions in the parameter space which allow
for the existence of such extrema.

\subsubsection{3- and k-dimensional cases}

For three neutrino species, we can write similarly to Eq.(\ref{e610})

\begin{equation}
f_m=\frac{B_1}{-m}+\frac{B_2}{i w_{12}-m}
+\frac{B_3}{i w_{13}-m}
=\frac{-1}{m} 
\frac{\left ( -a_1 + m\right ) \left ( -a_2+ m\right )} 
{\left ( -iw_{12} +m\right )\left ( -iw_{13}+m\right)}
\end{equation}

where $a_1,a_2$ are the roots of a certain 2-degree polynomial, they obey:

\begin{equation}
\left\{
\begin{array}{rl}
a_1+a_2&=i\left ( w_{12} (1-B_2)+w_{13} (1-B_3) \right )\\ \\
a_1 a_2&=-B_1 w_{12}w_{13}
\label{e200}
\end{array}
\right .
\end{equation}

and the product of them

\begin{equation}
\prod_{m=1}^{n-1} f_m=\frac{1}{B_1}\frac{(-1)^{n-1}}{ 
(n-1)!}
\frac{( -a_1)_{(n)} (-a_2)_{(n)}}{ (\beta_1)_{(n)}   (\beta_2)_{(n)} }
\end{equation}

with $\beta_{1,2}=-i w_{12,13} $. In the simplification of
the previous equation has been important the expression for $a_1 a_2$ in 
Eq.({\ref{e200}}). Note that this last expression is only valid if $w_{12}$ and
$w_{13}$ are both different from zero.

For the diagonal elements of U, we can write immediately, in terms of the generalized  Hypergeometric 
functions (Definition~(\ref{a140}))

\begin{equation}
<b \mid U\mid b>= e^{-i E_b (t-t_0)}\  {}_{2} F_2 (-a_1,-a_2,
-i w_{b,k1},-i w_{b,k2};z)
\end{equation}
with

\begin{equation}
\left\{
\begin{array}{rl}
a_1+a_2&=i\left ( w_{b,k1} (1-B_{k1})+w_{b,k2} (1-B_{k2}) \right )\\ \\
a_1 a_2&=-B_b w_{b,k1}w_{b,k2}
\end{array}
\right .
\label{e701}
\end{equation}
$k_1,k_2$ are labels for the two eigenvectors different
 from $b$.

For the $<b\mid U\mid a>$ ($a\ne b$), we can repeat exactly the same path as we 
did in the 2-dimensional section. 
We arrive to (we do it already for the general case):
\begin{equation}
z g'+ \beta_1 g= \frac{a_1\dots a_{k-1}}{\beta_1\dots \beta_{k-1}}z \ {}_{2}
F_{2}(-a_1+1,\dots,-a_{k-1}+1;\beta_1+1,\dots,\beta_{k-1}+1;z),\quad g(0)=0
\end{equation}

with $\beta_1=-iw_{ba},\beta_k=-iw_{bk}$, where $k\not=a,b$.
The solution of which is

\begin{eqnarray}
g(z) & = &\frac{a_1\dots
a_{k-1}}{\beta_1\dots\beta_{k-1}}z^{-\beta_1}\int_0^zdw w^{\beta_1}\  {}_{k-1} F_{k-1}
(-a_1+1,\dots,-a_{k-1}+1; \beta_1,\dots,\beta_{k-1} ; w ) \cr 
 & = &
\frac{a_1\dots a_{k-1}}{\beta_1\dots\beta_{k-1}}
z\int_0^1 ds s^{\beta_1}\ {}_{k-1} F_{k-1}
(-a_1+1,\dots,-a_{k-1}+1; \beta_1,\dots,\beta_{k-1} ; s z )\cr
&=&\frac{a_1\dots a_{k-1}}{\beta_1\dots\beta_{k-1}}
\frac{z}{1+\beta_1}\  {}_{k-1} F_{k-1}
(-a_1+1,\dots,-a_{k-1}+1; \beta_1+2,\beta_2+1,\dots,\beta_{k-1}+1 ;  z )\nonumber\\
\end{eqnarray}

Where we have used the  expression Eq.(\ref{a150}).
for $\mu=1$

   So
   
\begin{equation}
   <b\mid U\mid a>= 
   e^{-i (E_b t - E_a t_0)} V_{ba}
    \frac{ z}{1+i w_{ba}}\  {}_2 F_2(-a_1+1,-a_2+1;
   -i w_{b,k1}+2, -i w_{b,k2}+1; z )
   \end{equation}
$a_1,a_2$ fullfill the equations (\ref{e701}) with
    $k_1=a,k_2$
    are labels for the eigenvectors different to $b$ .

   Let's take for u 
   a general unitary 3-dimensional parametrized as follows

\begin{equation}
   u=\pmatrix{1 & 0 & 0 \cr
	      0 & C\psi & S\psi \cr
	      0 & -S\psi & C\psi \cr}
     \pmatrix{1 & 0 & 0 \cr
	      0 & e^{i\delta} & 0 \cr
	      0 & 0 & e^{-i\delta} \cr }
     \pmatrix{C\phi & 0 & S\phi \cr
	       0    & 1 & 0     \cr
	      -S\phi& 0 & C\phi \cr }
      \pmatrix{C\omega & S\omega & 0 \cr
	       -S\omega & C\omega & 0 \cr
		 0      &   0     & 1 \cr }
   \end{equation}

The potential V is
   $$ V_{ab}\equiv  u_{1b}^{\phantom{-1}}u_{a1}^{-1} =u_{1b} u_{1a}$$
The last equality is valid only in our particurlar case (or for any real u).

 $$u_{1a}(a=1-3)=(C\phi C\omega, C\phi S\omega, S\phi)$$

   Note that the phase $\delta$ and the angle $\psi$ don't appear in V.

In the basis where U is diagonal $U'= u U u^{-1} $.
   In particular, for the e-e transition we have
   $${\cal A}_{11}= U'_{11}=\sum_{kl} u_{1l} (u^{-1})_{k1}U_{lk}=\sum_{kl} V_{kl} U_{kl}=tr UV $$
In the small mixing limit   
the terms $V_{kl}$ are nearly zero for $k,l\ne 1$. Averaging out time dependent
terms,    we  write approximately:
   
\begin{equation}
   P_{ee}=\mid U_{11}'\mid^2\simeq C^2\phi C^2\omega \mid U_{11}(={}_2 F_2)\mid^2
   \end{equation}
with obvious arguments for the hypergeometric function.
The computation of a exact, fully simplified expression for $U_{11}'$ requires
the algebraic manipulation of several ${}_2 F_2$ functions and is not of major
interest. For the expected degree of approximation of this formula, see the 
comments correspondent to the 2-dimensional case.

   Now we are going to study the behaviour of $U_{11}$ for two different limits.
Let's suppose that the first and second eigenvalues are nearly degenerate:
   $w_{12}\to 0$, in this case

\begin{equation}
   \left\{
   \begin{array}{rcl}
   a_1 & = &  i w_{13} (1-B_3) \\
   a_2 & = & - B_1 w_{12} w_{13}\to 0 \\
   \end{array}
   \right .
   \end{equation}

   and 
   \begin{eqnarray}
   U_{11}&\propto&\ 1-\frac{a_2}{w_{12}}(\mbox{${{}_1 F_1}$}(-a_1,-i w_{13}; z)-1)
   \nonumber \\
    &=& 1-B_1 w_{13}(1-\mbox{${{}_1 F_1}$}(-i w_{13} (1-B_3),-i w_{13};z) 
   \end{eqnarray}
   The 3-oscillations has been reduced essentially to a 2 dimensional problem. We
   can apply to this last formula the  arguments of previous section for finding
   different limiting behaviours as function of the size of $w_{13}$.

   It is interesting also to study what happens when one of the eigenvalues
   is much bigger than the others (but none goes to zero) , so some of the $w$'s goes to infinity.
   Let's suppose $w_{13}\to\infty$. In such a limit
   , solving previously the second degree algebraic equation, we have

\begin{equation}
   \left\{
   \begin{array}{rcl}
   a_1 & = & \frac{ i w_{12} (1-B_2)}{2} \\
   a_2 & = &  i w_{13} (1-B_3) = i w_{13} (B_1+B_2) \\
   \end{array}
   \right .
   \end{equation}

   Thus
   \begin{eqnarray}
   \lim_{w_{13}\to \infty} U_{11} & \propto &\lim_{w_{13}\to\infty}\
   {}_2 F_2( \frac{i w_{12} (1-B_2)}{2}, i w_{13} (B_1+B_2), iw_{12}, i w_{13}; z)
   \nonumber \\
   & = &  \mbox{${{}_1 F_1}$}(\frac{i w_{12} (1-B_2)}{2},  iw_{12}; z
   (B_1+B_2))
   \label{e210}
   \end{eqnarray}
   so, as  expected, the system behaves effectively as a 2-dimensional system.
   Note that the 2-dimensional submatrix of u is not unitary, so the 
   sum $B_1+B_2$ is not equal to unity anymore in Eq.(\ref{e210}).
This procedure  is specially useful for the study of oscillations
in models with extremely heavy sterile neutrinos as those 
inspired in SO(10) GUT theories.

   For k dimensions, we can write similarly to Eq.(\ref{e610}) 
   
\begin{equation}
   f_m=\frac{B_1}{-m}+\frac{B_2}{i w_{12}-m}
   +\dots +\frac{B_k}{i w_{1k}-m}
   =\frac{(-1)^k}{m} 
   \frac{\left ( (-a_1) + m\right )\dots \left ( (-a_{k-1})+ m\right )} 
   {\left ( -iw_{12}+m\right )\dots\left ( -iw_{1k}+m\right)}
   \end{equation}

   where $a_1,\dots,a_{k-1}$ are the roots of the  (k-1)-degree 
   polynomial,
   \begin{equation}
   p(m)=\sum_{s=1}^{k} B_s \prod_{j\ne s}(m -i w_{1,j})
   \end{equation}

   The coefficient of the power of greatest degree, $c_0$, 
and the constant term, $c_{k-1}$ of this polynomial are:
   $$c_0=\sum_s B_s=1$$
   $$c_{k-1}=p(0)=\sum_{s=1}^{k} B_s \prod_{j\ne s}( -i) w_{1,j}=
   (-i)^{k-1} B_1 \prod_{j\ne 1} w_{1j} $$
    the product of the roots is then
   $$ a_1\cdot\dots \cdot a_{k-1}= (-1)^{k-1}\frac{c_{k-1}}{c_1}=
    i^{k-1} B_1 \prod_{j\ne 1} w_{1j} $$

Based in the previous formula 
and assuming all the $w_{ij}$ different from zero, we can generalize the 2 and 3-dimensional
expressions for the product,     
   
\begin{equation}
   \prod_{m=1}^{n-1} f_m=\frac{1}{B_1}\frac{(-1)^{n}}{
   (n-1)!}
   \frac{( -a_1)_{(n)}\dots (-a_{k-1})_{(n)}}{ (\beta_1)_{(n)}\dots   (\beta_{k-1})_{(n)} }
   \end{equation}

   with $\beta_{s}=-i w_{bs} $.

As it can be deduced trivially, the matrix U for k dimensions is 
the product
\begin{eqnarray}
U(t)&=&\exp-i H_0 (t-t_0) U_{red}( H^0,V, \rho_0,t_0)\nonumber
\end{eqnarray}
The elements of $U_{red}$ being essentially Generalized Confluent Hypergeometric
Functions of order k-1. The diagonal elements are given in a basis of eigenvalues of $H^0$
by:
   
\begin{equation}
   <a\mid U_{red} \mid a>=  {}_{k-1} F_{k-1} (-a_1,\dots,-a_{k-1},\beta_1,\dots,\beta_{k-1};z)
   \end{equation}
   with the $\alpha's$    and $\beta$'s 
   defined as before, and $z=-ie^{-t0/\lambda}\rho_0/\lambda$.
   
For non-diagonal elements:
   
\begin{equation}
   <b\mid U_{red}\mid a>= \frac{V_{ba}}{V_{bb}}\ \frac{z}{1+\beta_1}\ {}_{k-1} F_{k-1}(-a_1+1,\dots,-a_{k-1}+1;
   \beta_1+2,\dots, \beta_{k-1}+1; z )
   \end{equation}
Note that for each individual element $\beta$'s and $\alpha$'s are differently defined.

As it was pointed out before the previous expressions are valid only for the
completely non-degenerate case: all the $w_{ij}$ different from zero. The
general case will be treated elsewhere, it is easy to see that the order of the
hypergemetric functions appearing there will decrease in one unit for each of
the w's equal to zero.

The unitarity of U for any hermitian H induces a tower of identities between
the absolute values of the Hypergeometric Functions generalizing the expressions (\ref{e202b},\ref{e202}) 
found for the 2-dimensional case.

   \subsubsection{Evolution for finite time.}

     Until now we have considered the evolution of the system from a finite
     $t_0$ until infinite time. 
     Now, using  general properties of the evolution operator, we will compute 
     the evolution for arbitrary finite intervals.

Any U, for any intermediate time $t_1$, obeys the 
composition property:
     $$
     U(t,t_0)=U(t,t_1)U(t_1,t_0)
     $$
Taking the limit $t\to\infty$ in the above expression we arrive (taking $t_0=0$
for simplicity)
     $$
     U_\infty(t,0)=U_\infty(t,t_1)U(t_1,0)
     $$

     By the subindex $\infty$ we want to remark we have computed U as in the previous
     sections, using some limiting process.
So , as $U$ has an inverse,
\begin{eqnarray}
     U(t,0)&=&U_\infty^{-1}(\mu,t)U_\infty(\mu,0) \nonumber \\
     &=&U_\infty^{\dagger}(\mu,t)U_\infty(\mu,0)  \nonumber \\
     &=&U_{red}^\dagger (t_0=t) e^{i H_0 (\mu-t)} e^{-i H_0 \mu} U_{red}(t_0=0)
\nonumber \\
     &=&U_{red}^\dagger (t_0=t) e^{-i H_0 t} U_{red}(t_0=0) 
\label{e620}
\end{eqnarray}
Where we have effectuated some relabing to let clear the time dependence. $\mu$
,appearing at first as a new arbitrary parameter, disappears from the last
expression. $U_{red}$ is the matrix depending on the diverse parameters of the
system, including $t_0$ (compare with Eq.(\ref{e150})).

The matrix defined by
Eq.~(\ref{e620})
 satisfies by construction the initial condition $U(0,0)=1$. In addition we must
check it satisfies the  Equation (\ref{e201}): inserting the
last expression we obtained, and eliminating the common factor
$U_{red}(t_0=0)$, we get

\begin{eqnarray}
-i\partial_t e^{i H_0 t} U_{red}  
& =& e^{i H_0 t} U_{red} (H_0+ \rho(t) V) \nonumber \\
& & \nonumber \\
  e^{i H_0 t} H_0 U_{red} 
-i e^{i H_0 t}
\partial_t  
U_{red} 
& =& e^{i H_0 t} U_{red}  (H_0+ \rho(t) V) \nonumber \\
& & \nonumber \\
{ [} H_{0} , U_{red}  {]}  & = & 
 i \partial_t U_{red} + \rho(t) U_{red}    V 
\label{e630} 
\end{eqnarray}  
the dependence of $U_{red}$ on t  comes from the substitution
$t_0=t$.

If 

\begin{equation}
[ H_0,U_{red} (t) ] \approx 0
\end{equation}
for example if the difference of the eigenvalues of $H_0$ is relatively small,
then Eq.~(\ref{e630}) has the trivial solution:

\begin{equation}
U_{red} (t)=\exp -i V\int_0^t g(\mu) d\mu
\end{equation}
and the U matrix becomes

\begin{equation}
U(t,0)=\exp -i H_0 t\  U_{red} (t)=\exp -i H_0 t\, \exp -i V\int_0^t g(\mu) d\mu
\end{equation}
this is a highly non-trivial expression valid for arbitrary $H_0,V$ and $g(t)$.
Some pertubation approximation expansion in $\Delta E$ can be implemented for
better approximations.

In our particular case  it is easy to check that the
previous Eq.(\ref{e630})
is really fullfilled. It converts to a system of 
algebraic- functional relations between
 Hypergeometric functions.
For simplicity we take the 2-dimensional case. Eq.~(\ref{e630}) is 
equivalent to only two independent relations, with the same notation as in previous sections:

\begin{eqnarray}
\partial_z F & =& F\ V_{11}+ \mid V_{12}\mid^2  G \nonumber \\
F & =&(- V_{12}+\beta/z) G +  \partial_z G 
\end{eqnarray}

with the help of Eq.(\ref{a120}), we obtain instead two algebraic
equations that can be proved to be identities with  Eqs.(\ref{a160}).
We just prove in detail the first one. First we identify $V_{11}\equiv \alpha,
V_{12}^2=\alpha (1-\alpha)$, so
the first term is
\begin{eqnarray}
\partial_z F(\alpha \beta,\beta,z) & =&\alpha
\mbox{${{}_1 F_1}$}(\alpha\beta+1,\beta+1;z)
\end{eqnarray}
the second term is 
\begin{eqnarray}
 F\ V_{11}+ \mid V_{12}\mid^2  G & =&
\alpha \mbox{${{}_1 F_1}$}(\alpha\beta,\beta;z)+(1-\alpha)\left (
\mbox{${{}_1 F_1}$}(\alpha\beta,\beta;z)-\mbox{${{}_1 F_1}$}(\alpha\beta,\beta+1;z)\right ) \nonumber \\ 
& =&
 \mbox{${{}_1 F_1}$}(\alpha\beta,\beta;z)-(1-\alpha)\mbox{${{}_1 F_1}$}(\alpha\beta,\beta+1;z)
\end{eqnarray}
In the first line we have  used Eq.~(\ref{e202}). The last expression is
just the identity (\ref{a160c}).

In summary, we have proved that the evolution operator for any finite time is
\begin{eqnarray}
U(t,0)     &=&U_{red}^\dagger (t) e^{-i H_0 t} U_{red}(0) 
\end{eqnarray}
with $U_{red}$ given in  the previous sections. Expressions can be
given now for the transition probabilities in the mass or weak basis as we did
before. However the expressions are rather involved and not so illuminating. For
almost any practical purpose we can stick to the expressions given
in the $t\to\infty$ limit.

\section{Conclusions and further discussion.}

Exact expressions for the transition probabilities of neutrinos propagating radially
in the sun are given. These are valid for an arbitrary number of neutrino
species. The solution,  very compact in terms of Hypergeometric Functions,
is  very suitable for analytical studies of resonance and for 
systematic numerical approximations. The computational task of
computing numerically this kind of functions makes however very
impractical the actual use of them in concrete applications for the moment.

The perturbative expansion method used, that in this particular case gives
an exact solution, is well suited to produce approximate results for the same
matrix form of the potential (all eigenvalues but one equal to zero) but for
arbitrary form of $\rho(t)$. The consideration of a finite number of
terms in the expansion is expected not to be useful but yes  the summation for
all orders of appropriate leading terms.
Another possibility is to consider the difference between an arbitrary
$\rho(t)$ and  our exponential as  small and do perturbation theory around our
exact solution. In principle we expect this procedure to be convenient at far
distances from the creation point, as we have seen the probability
transition depends more on the existence of a resonant region  than on 
the  detailed
shape of the potential. This would make possible to compute formulas for
non-radial propagation and for the regions where the exponentially decaying
density lose validity: central core and solar surface.

The phenomenon of resonance (or MSW effect) is remarkable in itself. The present
solution offers a starting point for its analytical description. We note the
similarity of this effect with the  original form of the Anderson Localization
effect presented in \cite{and1}.
This similarity is specially evident  under the treatment done in this work.
In both cases  the  locking of a system in one of its quantum states
except for a certain range of the parameter space arise as a property
of the solution of a coupled system of ordinary differential equations. As in its
case (which physically correspond to the evolution of a particular
site in a random lattice) the proper energies (or its differences) of each mode play an important
role, however here these are not needed to be stochatically distributed. 
This condition of randomness is essential in \cite{and1}, the results in this
work
induce us to think whether such randomness is a mere
mathematical convenience more than having a deep physical meaning.

Apart from physical interest, our solution has intrinsic mathematical interest
in the Theory of Differential equations and of Special Functions.
A general theory of equations of the form
$$\partial_t x= (A+f(t) B) x $$
with A,B constant matrix is missing in spite of the fact these systems are
the ''next-step'' in complexity from all-constant coefficient equations. Only
solutions for a few examples of these equations are known.

 The
generalized Hypergeometric functions are shown to be the asymtotic solutions of
a simple difererential equation in a systematic way. New identities for the
absolute values of these functions are derived. The generality of the results
obtained here induces us to consider the utility of defining Hypergeometric
functions of matrix parameters $\mbox{${{}_1 F_1}$}(A,B;z)$ with z complex, similar extensions
exist already: generalizations
 with complex parameters but matrix argument or parameters and argument defined
in arbitrary finite fields  for example.
Let's note finally that making the change   
$$y=\exp(-t)$$
our system becomes of the form
$$\partial_y x=\left (\frac{A}{y}+B \right ) x$$
so we have found also a solution for this
system, the Hypergeometric functions appear now as regular
solutions for  $y\to 0$.

  \appendix

  \section{Appendix: some formulas about Hypergeometric Functions}

  See \cite{grad} and references therein for all of these definitions and formulas.
   The Confluent Hypergeometric function  is defined by
  
\begin{equation}
  \mbox{${{}_1 F_1}$}(a,b;z)=\sum_{n=0}^\infty \frac{(a)_{(n)}}{(b)_{(n)}} \frac{z^n}{n!}
  \label{a100}
  \end{equation}
where    the Pochammer symbol is 
  
\begin{equation}
  (z)_{(n)}=\Gamma(n+z)/\Gamma(z)
  \label{a110}
  \end{equation}

  The generalized Hypergeometric function is defined by
  
\begin{equation}
  {}_p F_q (a_1,\dots,a_p,b_1,\dots,b_q; z)=\sum_{n=0}^\infty 
  \frac{(a_1)_{(n)}\dots (a_p)_{(n)}}{(b_1)_{(n)}\dots (b_q)_{(n)} }
    \frac{z^n}{n!}
  \label{a140}
  \end{equation}

  The integral

\begin{equation}
  \int_0^t x^{\gamma-1} (t-x)^{c-\gamma-1} \mbox{${{}_1 F_1}$}(a,\gamma; x)dx=
  t^{c-1}\frac{\Gamma(\gamma)\Gamma(c-\gamma)}{\Gamma(c)}\ {}_1 F_1(a,c;t);
  \quad [ \Re\ c,\gamma >0]
  \label{a130}
  \end{equation}

is a special case of

  \begin{eqnarray}
  \lefteqn{\int_0^1 (1-x)^{(\mu-1)} x^{b_1-1} {}_p F_q
  (a_1,\dots,a_p;b_1,\dots,b_p; ax)dx =} \nonumber\\
  & & \frac{\Gamma(\mu)\Gamma(b_1)}{\Gamma(\mu+b_1)} {}_p F_q(a_1,\dots,a_p;\mu+b_1,b_2,\dots,b_q;a); \nonumber\\
  & & \quad \left (\Re(\mu,b_1)>0,\ p<q+1 \right )
  \label{a150}
  \end{eqnarray}

Some other important formulas are:
\begin{eqnarray}
    \frac{d\ \mbox{${{}_1 F_1}$}(\alpha,\gamma;z)}{dz}&=&\frac{\alpha}{\gamma}\
  \mbox{${{}_1 F_1}$}(1+\alpha,1+\gamma;z)  \label{a120} \\
\mbox{${{}_1 F_1}$}(\alpha,\gamma; z)&=& \exp z\  \mbox{${{}_1 F_1}$} (\gamma-\alpha,\gamma; -z) \label{a121}\\
\frac{z}{\gamma}\mbox{${{}_1 F_1}$}(\alpha+1,\gamma+1; z)&=&  \mbox{${{}_1 F_1}$} (1+\alpha,\gamma; z)-
\mbox{${{}_1 F_1}$}(\alpha,\gamma;z) \\
\alpha \mbox{${{}_1 F_1}$}(\alpha+1,\gamma+1; z)&=&(\alpha-\gamma)  \mbox{${{}_1 F_1}$} (\alpha,\gamma+1;
z)+\gamma \mbox{${{}_1 F_1}$}(\alpha,\gamma;z)\label{a160c} \\
\alpha \mbox{${{}_1 F_1}$}(\alpha+1,\gamma; z)&=&(z+2\alpha-\gamma)  \mbox{${{}_1 F_1}$} (\alpha,\gamma;
z)+(\gamma-\alpha) \mbox{${{}_1 F_1}$}(\alpha-1,\gamma;z) 
\label{a160}
\end{eqnarray}

From \cite{hyzero} we know that the real zeros $x_j$ of 
$\mbox{${{}_1 F_1}$} (a,c;x)$
for a,c real satisfy the bounds

\begin{equation}
 (c-2a)-2\surd (a(a-c)-c )< x_j < (c-2a)+2\surd (a(a-c)-c) 
\label{a170}
\end{equation}
 The smallest  real zero $x_1=x_{min}$ satisfy

\begin{equation}
 x_{min} < \frac{ c(c+2)}{c-2a}
\label{a180} 
\end{equation}

Appliying the same bound to the expression (\ref{a121})
we deduce 
the lower bound for the maximal real zero $x_{max}$

\begin{equation}
( x_{min} <) \frac{ c(c+2)}{c-2a} < x_{max} 
\label{a190}
\end{equation}
\vspace{1cm}
{\Large{\bf Acknowledgements.}}

I would like to thank to Peter Minkowski for many enlighthening discussions. This 
work has been supported in part by the Wolffman-Nageli Foundation (Switzerland)
and by the MEC-CYCIT (Spain).

\newpage

\newpage
\begin{thebibliography}{99}

\bibitem{panta1} T.Kuo,J.Pantaleone. {\em Phys. Rev. Lett. } 57,14 (1986)
p.1805.
 PURD-TH-89-3 (1989).



\bibitem{mik1} S.P. Mikheyev, A. Smirnov. {\em Phys. Lett. B}. Vol.200,4
(1988) p.560.
 {\em Il Nuovo  Cimento.}. Vol.9 (1986) p.17.

\bibitem{pet1} S. T.  Petcov. {\em Phys. Lett. B} vol.214,2 (1988) p.259.

\bibitem{bah1} J.N. Bahcall, R. Ulrich. {\em Rev. Mod. Phys.} 60 (1988) 267.






\bibitem{zen1} C. Zener. {\em Proc. Royal Soc. London}. Ser A,137 (1932)
p.696.

 L. Landau. {\em Phys. Z. Soviet Union} 2-46 (1932).

\bibitem{pet2} S. T.  Petcov. {\em Phys. Lett. B} Vol.191 (1987) p.299.

\bibitem{hax1} W.C. Haxton, {\em Phys. Rev. D} 35 (1987) 2352.

\bibitem{not1} D. Notzold, MPI-PAE/Pth 08/87 (Munich 87).

\bibitem{pet3} S. T.  Petcov. {\em Phys. Lett. B} vol.200,3 (1988) p.373.

\bibitem{tov1} U. Toshev. {\em Phys. Lett. B196} (1987) 170.

\bibitem{aba1} A. Abada, S.T.  Petcov. {\em Phys. Lett. B} Vol.214,2 (1988)
p.139. {\em Phys. Lett. B} vol.279 (1992) p.153.

\bibitem{grad} I.S. Gradshteyn, I.M.Ryzhik. {\em Table of Integrals,series and
products}. Academic Press.1980.

\bibitem{din1} R.B.Dingle. {\em Asymtotic Expansions}. Academic Press.1973.

\bibitem{panta2} T.Kuo,J.Pantaleone. 
 {\em Phys. Rev.} {\bf D35,11} (1987) p.3432.





\bibitem{and1} P.W. Anderson. ''Absence of Diffusion in Certain Random
Lattices''. {\em Phys. Rev.} {\bf 109-5}, 1492-1507 (1958).
\bibitem{hyzero} S. Ahmed {\em J. of Approximation Theory} {\bf 34}, 335-347 (1982).
\end{thebibliography}
\end{document}